\documentclass{sf2a-conf2017}
\usepackage{graphicx}
\usepackage{hyperref}
\usepackage[]{natbib}  
\usepackage{epstopdf}

\def\BibTeX{{\rm B\kern-.05em{\sc i\kern-.025em b}\kern-.08em
    T\kern-.1667em\lower.7ex\hbox{E}\kern-.125emX}}
\bibpunct{(}{)}{;}{a}{}{,}  %%%%%%%%%%%%%  A&A bibliography style
\newcommand{\powvar}{{FliPer} }
\newcommand{\powvarp}{{FliPer}}
\newcommand{\numax}{{$\nu_{\rm{max}}$} }
\newcommand{\numaxp}{{$\nu_{\rm{max}}$}}

\newcommand{\az}{{A2Z} }
\newcommand{\azp}{{A2Z}}
\newcommand{\kepler}{{\emph{Kepler}} }
\newcommand{\keplerp}{{\emph{Kepler}}}
\newcommand{\kto}{{K2} }
\newcommand{\ktop}{K2}
\newcommand{\dnu}{{$\Delta \nu$}}

\begin{document}

\TitreGlobal{SF2A 2017}

%%-----------------------------------------------------------------
%%      the top matter
%%

\title{FliPer: checking the reliability of global seismic parameters from automatic pipelines}%A new metric to estimate the surface gravity of stars}% from Main sequence stars to Red Giants}% study global seismic parameters}

\runningtitle{FliPer method}

\author{L. Bugnet$^{1,}$}\address{IRFU, CEA, Universit\'e Paris-Saclay, F-91191 Gif-sur-Yvette, France}\address{Universit\'e Paris Diderot, AIM, Sorbonne Paris Cit\'e, CEA, CNRS, F-91191 Gif-sur-Yvette, France}
\author{R. A. Garc{\'{\i}}a$^{1,2}$}
\author{G. R. Davies$^{3,}$}\address{School of Physics and Astronomy, University of Birmingham, Edgbaston, Birmingham, B15 2TT, UK}\address{Stellar Astrophysics Centre, Department of Physics and Astronomy, Aarhus University, Ny Munkegade 120, DK-8000 Aarhus C, Denmark}

%% IF Author3 has the same affiliation than Author1:

%% IF Author3 has its own affiliation:
%\author{C.\,E. Author3}\address{Dept. of Chess, University of Games, 35101 Las Vegas, Monaco} 

%% IF Author3 has two affiliations, the one of Author1 and a second one:
%\author{C.\,E. Author3$^{1,}$}\address{Dept. of Chess, University of Games, 35101 Las Vegas, Monaco} 

\author{S. Mathur$^{5,6,}$}\address{Instituto de Astrof\'{\i}sica de Canarias, E-38200, La Laguna, Tenerife, Spain}\address{Universidad de La Laguna, Dpto. de Astrof\'{\i}sica, E-38205, La Laguna, Tenerife, Spain}\address{Space Science Institute, 4750 Walnut Street Suite 205, Boulder, CO 80301, USA}

\author{E. Corsaro}\address{INAF - Osservatorio Astrofisico di Catania, Via S. Sofia 78, I-95123 Catania, Italy}
%% Keep this line, even if the page will be settled afterwards.
\setcounter{page}{1}

%%-----------------------------------------------------------------

\maketitle
% * <grd349@gmail.com> 2017-10-20T07:54:26.027Z:
% 
% Title is a bit long.  How about "FliPer: Fault finding for Asteroseismic pipelines "
% 
% ^.
%%-----------------------------------------------------------------
%%        The abstract
%% 
%%  Warning!  within the abstract:
%%  - do not use macros. 
%%  - do not use commands like: \cite, \citet, \citep ... etc.

\begin{abstract}
Our understanding of stars through asteroseismic data analysis is limited by our ability to take advantage of the huge amount of observed stars provided by space missions such as CoRoT, \keplerp, \ktop, and soon TESS and PLATO. Global seismic pipelines provide global stellar parameters such as mass and radius using the mean seismic parameters, as well as the effective temperature. These pipelines are commonly used automatically on thousands of stars observed by K2 for 3 months (and soon TESS for at least $\sim$ 1 month).  However, pipelines are not immune from misidentifying noise peaks and stellar oscillations. Therefore, new validation techniques are required to assess the quality of these results. We present a new metric called FliPer (Flicker in Power), which takes into account the average variability at all measured time scales. The proper calibration of \powvar enables us to obtain good estimations of global stellar parameters such as surface gravity that are robust against the influence of noise peaks and hence are an excellent way to find faults in asteroseismic pipelines.

\end{abstract}

%% Insert the keywords (to appear in the ADS indexing)
%% Keywords must be separated by a comma

%%-----------------------------------------------------------------

\section{Introduction}
%%---------------------
%Our understanding of stars through the study of their global properties such as mass and radius is limited by the accuracy on the value of the surface gravity we estimate from observatiions. 
Surface gravity and global seismic parameters (\dnu, \numaxp) are related through the so-called global seismic scaling relations (\cite{Brown}, \cite{Kjeldsen}):
%\begin{equation}
         ${\Delta \nu} \propto {M}^{\frac{1}{2}} \times {R}^{\frac{-3}{2}} $
         %\label{echelle1}
%\end{equation}
%\begin{equation}
		and ${\nu_{max}} \propto {M} \times {R}^{-2} \times {T_{eff}}^{-\frac{1}{2}}$. Hence, an accurate estimation of the seismic global parameters and the effective temperature can be used to provide an estimate of the surface gravity of stars with convective envelopes. However, most asteroseismic data obtained from \kepler and \kto are sampled with cadence of around 30 minutes (long cadence), leading to limited spectral information above the corresponding Nyquist frequency ($\sim$ 288 $\mu$Hz). If a star pulsates at frequencies higher than the long cadence Nyquist frequency (for instance a main-sequence star, e.g. \cite{Davies15}) typical analysis methods cannot be applied to estimate seismic parameters. In some cases reliable information can be obtained (\cite{Chaplin}) but typical automated asteroseismic pipelines are susceptible to providing unreliable estimates. Indeed, internal magnetic fields can inhibit the modes (e.g. \cite{Mosser09}, \cite{Garcia10}, \cite{Chaplin11}), complicating the automatic characterization of seismic parameters.
% * <grd349@gmail.com> 2017-10-20T08:09:20.529Z:
% 
% > Even in the range of visible frequencies, magnetic fields can inhibit the modes (e.g. \cite{Mosser09}, \cite{Garcia10}, \cite{Chaplin11}), complicating the characterization of seismic parameters.\\
% 
% This seems to come out of the blue some what ... This probably deserves its own paragraph or if you are short on space then leave it for the main paper.
% 
% ^.
% * <grd349@gmail.com> 2017-10-20T08:03:13.743Z:
% 
% After the scaling relation references (Brown, Kjeldsen), write out the scaling relation.
% 
% ^ <grd349@gmail.com> 2017-10-20T14:51:08.836Z.

Several methods have been developed to estimate the stellar parameters (e.g. surface gravity) from photometric data based on a simple measurement such as the variance of the time series (\cite{Hekker12}), the Flicker (\cite{Bastien13}, \cite{Bastien16}), granulation characterization of the power spectrum (\cite{Mathur11}, \cite{Kallinger14}) or the autocorrelation as described in \cite{Kallinger16}. Some of these methods require the observation of acoustic modes of oscillation but those that rely on the information provided by just granulation do not. The Flicker technique is typically used for main-sequence stars, subgiants and giants down to a $\log_{10}(g)$ $\sim$2.5 dex. %INSERT FLICKER SAVITA %(\cite{Kallinger}, \cite{mathur], \cite{bastien}). 
% * <grd349@gmail.com> 2017-10-20T08:12:46.707Z:
% 
% > The Flicker is mostly useful for main-sequence stars, subgiants and giants down to a log(g) $\sim$2.5. 
% 
% This statement does not tie up with the thrust of the paper - For the checking of seismic pipelines and identifying of supernyquist stars, the giants are the most useful. 
% 
% ^ <grd349@gmail.com> 2017-10-20T08:14:46.496Z:
% 
% OK - I see what you are doing.  I have changed this to make it clearer.
%
% ^.
% * <grd349@gmail.com> 2017-10-20T08:10:47.736Z:
% 
% > While the latter requires the observation of acoustic modes, the former does not have such requirement.
% Change to : Some of these methods require the observation of acoustic modes of oscillation but those that rely on the information provided by just granulation do not.
% 
% ^.
With FliPer we can reach the range of application to red giants to lower log(g) by taking into account the variability at all measured frequencies in the power spectrum (Bugnet et al. \textit{in prep.}). Thus, with this simple metric it is possible to assess in a few seconds the reliability of results obtained from automatic global seismic pipelines.
% * <grd349@gmail.com> 2017-10-20T08:15:36.912Z:
% 
% > With FliPer we can extend the range of application to red giants to lower log(g) by taking into account the variability at all measured frequencies in the power spectrum
% 
% Bastien+ already have the red giant flicker in there already.  I would avoid saying 'we can extend'.
% 
% ^.

\begin{figure}[ht!]
\centering\includegraphics[width=0.97\hsize, clip]{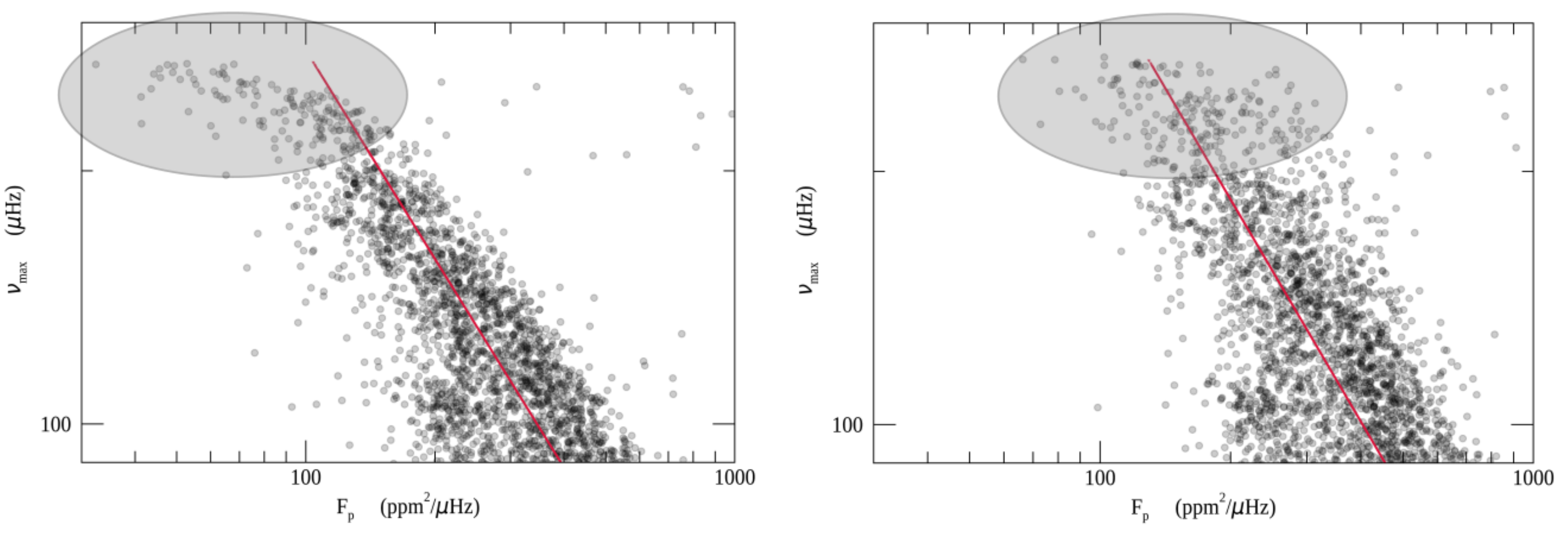}
\caption{\numax provided by the \az pipeline (\cite{Mathur}) \textbf{vs} \powvar for stars with \numax from $90$ to $300 \mu Hz$ (It corresponds to a zoom on the rectangle on the left panel of Fig.~\ref{RG}). \textbf{Left:} Case where \powvar is computed using the mean of the high-frequency signal as the photon noise. \textbf{Right:} Case when the theoretical photon noise computed by \cite{Jenkins} is used. The grey-shaded circle is where the impact of the noise calculation on \powvar is important.}
\label{ex}
\end{figure}

\section{Origin and definition of FliPer}
%%-------------------------
It is well known that the shape and the amount of power in the power spectrum density (PSD) of a star changes with stellar evolution as discussed above. The shape of the PSD is dominated by photometric variability caused by star spots and stellar rotation at low frequencies, a convective continuum, and a hump of power due to the stellar oscillations. All of these stellar contributions are added to the constant photon noise. As a star evolves and surface gravity decreases, so \numax decreases.  Because granulation properties are linked to \numax or surface gravity, as a star evolves so the total power from granulation increases (e.g. \cite{Garcia}).
%The properties of the granulation and the oscillations combined yield to a shifting of \numax to low frequencies and an increase of the amount of power as the star evolves (e.g. \cite{Garcia}). 
To account for the total power in the PSD we define the new metric FliPer ($F_{p}$) as follows:
% * <grd349@gmail.com> 2017-10-20T08:19:30.955Z:
% 
% > (Ref, ref, ref)
% Either add refs or say as discussed above.
% 
% ^.
% * <grd349@gmail.com> 2017-10-20T08:17:33.390Z:
% 
% > The properties of the granulation and the oscillations combined yield to a shifting of \numax to low frequencies and an increase of the amount of power as the star evolves (e.g. \cite{Garcia}).
% 
% Change this to ... As a star evolves and surface gravity decreases, so \numax decreases.  Because granulation properties are linked to \numax or surface gravity, as a star evolves so the total power from granulation increases. ... or something like that.
% 
% ^.
\begin{equation}
F_p= \overline{\textsc{PSD}} -	P_n
\end{equation}
where $\overline{\textsc{PSD}}$ represents the mean value of the power spectrum density and $P_n$ the photon noise. $\overline{\textsc{PSD}}$ is computed from 0.7 $\mu$Hz (corresponding to the 20 days high-pass filter used to calibrate the light curves following \cite{Garcia11}) to the Nyquist frequency ($\sim$288 $\mu$Hz for long cadence \kepler data). The photon noise is estimated following \cite{Jenkins} and depends on the magnitude of the star.
% * <grd349@gmail.com> 2017-10-20T08:42:34.862Z:
% 
% > Noise estimation}
% > \label{sec:noise}
% 
% This section should be higher up the document in the methods section, i.e., before the results.
% 
% ^ <grd349@gmail.com> 2017-10-20T08:44:36.989Z:
% 
% It also needs to have more information included to help the reader understand.  
%
% ^ <grd349@gmail.com> 2017-10-20T15:13:03.307Z.
It could also be evaluated by taking the mean power at high frequency (see Fig. \ref{ex} left panel) instead of computing the expected noise by \cite{Jenkins} (see Fig. \ref{ex} right panel). A comparative study showed that for most stars, the values of \powvar obtained with both methods are similar. The only important difference appears for stars in which the high-frequency part of the spectrum is dominated by stellar signal and not by noise. This is typically the case for stars with \numax close to Nyquist or for super-Nyquist stars. For these stars, the power contained in the modes and in the granulation profile is partially taken into account in the noise calculation because there are still power at high frequencies. The resulting photon noise value is thus higher than expected. In this case, \powvar is artificially lowered when using the high-frequency noise calculation as shown for stars in the grey circle on Fig.~\ref{ex}.\\

\begin{figure}[ht!]
\centering
\includegraphics[width=\hsize, clip]{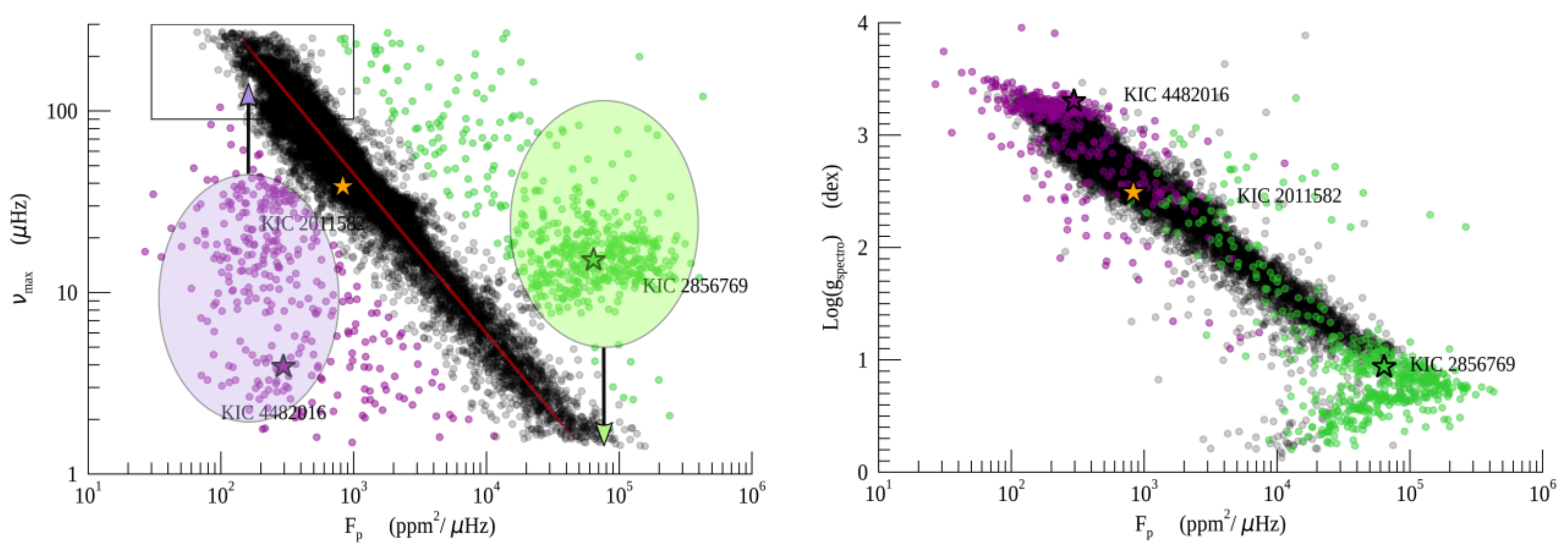}
\caption{\textbf{Left:} \numax provided by the \az pipeline (\cite{Mathur}) for $\sim 16,000$ red giants observed by \kepler (data corrected and interpolated following \cite{Garcia11,Garcia14c})$\,$ \textsc{vs} \powvarp. The red line represents the best fitting to the data. %The pink shaded area represent the $1 \sigma$ distance from the law. 
Green stars have higher \powvar than the general trend and should be replaced at lower frequencies, while violet points have lower values of \powvarp and should be replaced at higher frequencies (as indicated by ellipses and arrows). These limits are located at one standard deviation from the law. The black rectangle shows the location of the zoom of Fig.~\ref{ex}. \textbf{Right:} $\log(g)$ provided by photometric and spectroscopic measurements from the \textsc{nasa} \kepler catalog \cite{Mathur17} \textsc{vs} \powvar for the same stars than in the left panel.}
% * <grd349@gmail.com> 2017-10-20T08:24:26.544Z:
% 
% > Green stars have higher \powvar than the general trend, while violet points have lower values of \powvarp. \textbf{Right:} $\log(g)$ provided by photometric and spectroscopic measurements from the \textsc{nasa} \kepler catalog \cite{Mathur17} \textsc{vs} \powvar for the same stars than in the left panel.
% 
% Say why the green and purple stars are there - i.e., super Nyquist etc ... 
% 
% ^.
\label{RG}
\end{figure}

\section{Prediction of seismic parameters}

From Fig.~\ref{RG} left, we get in a first approximation a logarithmic law ($\log(F_{p, ppm^2/\mu Hz})=-1.14\times \log(\nu_{max, \mu Hz}) + 4.88$) between \powvar and \numax followed by more than 90\% of the $16, 000$ red giants analyzed here. By calculating \powvarp, we used this metric to determine those stars for which the resultant seismic parameters do not follow the general trend. This could be a consequence of the presence of unexpected features in the PSD (e.g. pollution by spikes, etc) or because the resultant value obtained by the pipeline is incorrect.
% * <grd349@gmail.com> 2017-10-20T08:26:44.397Z:
% 
% >  we have used this 
% Try to write in the past tense (it's a good general rule that if you obey it, makes writing easier).
% 
% ^ <grd349@gmail.com> 2017-10-20T15:13:14.611Z.
% * <grd349@gmail.com> 2017-10-20T08:25:32.553Z:
% 
% > From Fig.~\ref{RG} \textbf{Left}, we get in a first approximation a logarithmic law (see equation \ref{powlaw}) between \powvar and \numax followed by more than 90\% of the $16, 000$ red giants analyzed here within 1-$\sigma$.
% 
% I'm not sure what 1-sigma means here (Normally if you take all X within 1 sigma you should have 68%.  More than 90% is closer to 2 sigma).
% 
% ^.

% \begin{equation}
% % * <grd349@gmail.com> 2017-10-20T08:28:32.593Z:
% % 
% % > \begin{equation}
% % > \log(F_{p})=-1.14\times \log(\nu_{max}) + 4.88
% % > \label{powlaw}
% % > \end{equation}
% % 
% % Do you have uncertainties on these values?
% % NO I DON'T HAVE ANY BECAUSE IT DEPENDS ON WHERE WE PUT THE LIMIT BETWEEN GOOD STARS AND OUTLIERS... FOR NOW THE LIMIT IS "RANDOM", I CHOSE THE DISTANCE FROM THE LAW THAT SEEMED TO MEAN SOMETHING.
% % ^.
% \log(F_{p})=-1.14\times \log(\nu_{max}) + 4.88
% \label{powlaw}
% \end{equation}

The right panel of Fig.~\ref{RG} represents the spectroscopic surface gravity from the \textsc{nasa} \kepler catalog (\cite{Mathur17}) against \powvar for the same sample of stars using the same color code than in the left panel. We observe that the purple outliers in the left panel follow the general trend with log(g), while the green data points appear to be reflected at some point.
% * <grd349@gmail.com> 2017-10-20T08:30:14.492Z:
% 
% > re still not all correct.
% You should expand on this.  Why is it that you say they are not correct?    Do you mean that the relation is reflected around the Nyquist?  
% 
% ^ <grd349@gmail.com> 2017-10-20T08:32:10.310Z:
% 
% Never mind - I see.
%
% ^.
This change of slope around 0.7 dex comes from the cut in the PSD as a consequence of the high-pass filter used to calibrate the data (\cite{Garcia11}). All stars with $\log(g)$ lower than this boundary have a biased estimation of \powvar and form a clump of green outliers stars on the left panel. This left panel show the same data from FliPer but with seismic \numax on the y-axis. The $\log(g)$ measurements come from spectroscopic analysis that are independent of the seismic analysis. We thus demonstrate that most outliers stars in the left panel are due to a problem in the automatic seismic determination and not in \powvarp, because \powvar values are consistent with surface gravity data. 
% * <grd349@gmail.com> 2017-10-20T08:35:24.974Z:
% 
% > In the left hand panel ...  
% 
% You jump from the right panel to the left panel too quickly.  You need to introduce the left panel first with something factual (The left panel show the same data from FliPer but now with seismic blah on the y-axis.  Then provide some interpretation).
% 
% ^.
% * <grd349@gmail.com> 2017-10-20T08:33:41.665Z:
% 
% > Because the log(g) measurements come from spectroscopic analysis that are independent of the seismic analysis, we demonstrate that the outliers stars in the left panel are due to a problem in the automatic seismic determination and not in \powvarp.
% 
% Is that true?  Haven't the spectroscopic log's been iterated with seismic info used to get the correct logg?
% 
% ^.

\begin{figure}[ht!]
\centering
\includegraphics[width=\hsize, clip]{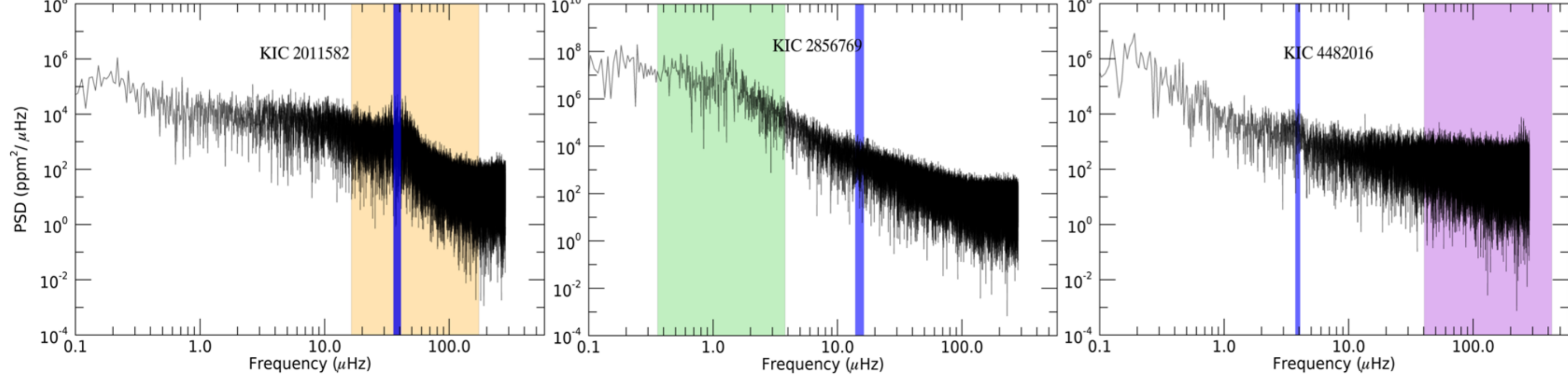}
\caption{PSD of the three stars represented with a star symbol on Fig.~\ref{RG}. The blue area corresponds to the \numax returned by \azp. The yellow, green, and purple shaded regions correspond to the range of accepted values of \numax by \powvarp as in Fig. \ref{RG}. \textbf{Left:} KIC~2011582 for which the \numax is well determined by \azp. \textbf{Middle:} A high F$_p$ star (KIC~2856769). This star has lower frequency modes than obtained by \azp. \textbf{Right:} A low F$_p$ star (KIC~4482016). This star has higher frequency modes than obtained by \azp.}
\label{spectre}
\end{figure}

Figure~\ref{spectre} represents the PSD of three stars represented with a star symbol in Fig.~\ref{RG}. Left panel corresponds to KIC~2011582 well characterized both by \az and \powvarp. Stars with a higher \powvar value than expected (green stars on Fig.~\ref{RG}) have low surface gravities, meaning that they are highly evolved RGB stars. The \numax is too close to the frequency cut-off of 20 days used to calibrate the series.  As a result, the A2Z pipeline cannot properly estimate \numax (a lower filter is needed to properly analyze these stars). An example, KIC~2856769, is presented in the middle panel in Fig.~\ref{spectre}. Most outliers stars with a low value of \powvar (purple stars on Fig.~\ref{RG}) have a high $\log(g)$: they are probably main-sequence or sub-giants stars and should have a much higher \numax than the value returned automatically by \azp. An example is KIC ~4482016 represented on the right panel of Fig.~\ref{spectre}. These stars needs to be treated independently by A2Z and \powvar helps to flag them up.\\

There are however about $1\%$ of the stars that remain outliers on the right panel of Fig. \ref{RG}. Among these outliers that could present however a good estimation of \numaxp, we observe stars that present high rotation power (e.g. \cite{Garcia14b}, \cite{Ceillier}), spikes, pollution by another star, binaries systems, low signal-to-noise ratio stars (\cite{Mathur16}), low-amplitude dipole mode stars (e.g. \cite{Mosser12},\cite{Garcia14a}), etc. Not only does the \powvar metric allows to estimate surface gravity, but also to detect stars that present a particular signal in their power spectra. For example, the detection of spikes is important in the study of \kto observations which are affected by spikes at the Thrusters frequency and its harmonics.

\section{Conclusions}
%%--------------------
We demonstrate that the \powvar follows a quasi-logarithmic trend with the global seismic parameters and, therefore, it is related to surface gravity. It allows us to quickly estimate the reliability of seismic parameters estimated from global pipelines. The \powvar method can be used to identify stars without detected modes, stars dominated by the harmonics of the K2 Thrusters seen as spikes in the spectrum, highly evolved stars, and super-Nyquist stars (i.e., stars for which the p-mode excess power is above the observational Nyquist frequency).
% Optional acknowledgements
% -------------------------
\begin{acknowledgements}
%The standard acknowledgement, if required, is : Thank you!
L.B. and R.A.G. acknowledge the support from CNES.
S.M. acknowledge support by the National Aeronautics and Space Administration under Grant NNX14AB92G issued through the \kepler Participating Scientist Program. E.C. is funded by the European Union’s Horizon 2020 research and innovation programme under the Marie Sklodowska-Curie grant agreement no. 664931.
\end{acknowledgements}

%%-----------------------------
%%   Bibliography
%%-----------------------------
%%
%% The reference list should contain all the references cited in the text, ordered alphabetically by surname (with
%% initials following). If there are several references to the same first author, they should be entered according
%% to the following scheme:
%% 1. One author: chronologically
%% 2. Author, one co-author: alphabetically by co-author, then chronologically
%% 3. Author, two or more co-authors: chronologically.
%%
%% Please note that for papers that have more than five authors, only the first three should be given, followed
%% by "et al."
%%
%% The format for references is the one adopted by A&A (see the example below).
%%
%% To set the reference list in the proper A&A format, we encourage you to use BibTEX and the natbib
%% package instead of the standard 'thebibliography' environment.
%%

% %% The following lines are required when using BibTEX (strongly encouraged!):
% \bibliographystyle{aa}  % A&A bibliography style file (aa.bst)
% \bibliography{sf2a-template} % your references in file: Yourfile.bib

%
\end{document}